**Direct fiber-coupled soliton microcomb system with enhanced stability and reproducibility via high numerical-aperture polarization-maintaining single-mode fibers and temperature control**

Miezel Talara,[1,†] Yu Tokizane,[1,†] Kodai Yamaji,[2] Yoshihiro Makimoto,[1,3] Kenji Nishimoto,[1,2] Yudai Matsumura,[2] Masayuki Higaki,[2] Naoya Kuse,[1,4] and Takeshi Yasui[1,4,*]

[1]Institute of Post-LED Photonics (pLED), Tokushima University, 2-1, Minami-Josanjima, Tokushima, Tokushima, 770-8506, Japan
[2]Graduate School of Sciences and Technology for Innovation, Tokushima University, 2-1, Minami-Josanjima, Tokushima, Tokushima 770-8506, Japan
[3]Tokushima Prefectural Industrial Technology Center, 11-2, Nishibari, Saika-Cho, Tokushima, Tokushima 770-8021, Japan
[4]Institute of Photonics and Human Health Frontier (IPHF), Tokushima University, 2-1, Minami-Josanjima, Tokushima, Tokushima, 770-8506, Japan
[†]These authors contributed equally.
*yasui.takeshi@tokushima-u.ac.jp

**Abstract**

We propose a compact and robust system architecture for soliton microcomb generation, based on two key techniques: direct fiber coupling using high numerical-aperture polarization-maintaining single-mode fibers (high-NA PMFs) and active temperature control of the microresonator. These complementary strategies address two major challenges in microcomb implementation: environmental sensitivity and resonance instability. Building on prior work using single-mode fiber (SMF)-based direct coupling, which demonstrated device miniaturization and partial suppression of thermal drift in coupling efficiency, our PMF-based approach offers enhanced thermal stability and significantly greater robustness to environmental disturbances such as temperature fluctuations and vibration. In our system, precision alignment using microscopes or multi-axis stages is no longer required, enabling a simplified optical setup and stable long-term operation. The direct coupling scheme achieved a coupling efficiency of 57.7% and maintained soliton operation for over 24 hours under external perturbations. In parallel, active temperature control of the microresonator was quantitatively evaluated, reducing the wavelength variation of the pump and auxiliary lasers by 79% and 97%, respectively. This stability enables reproducible soliton generation even in thermally dynamic environments. Comparative experiments with SMF-based direct coupling and lensed-SMF-based free-space coupling systems confirmed the superior performance of the PMF-based design in terms of coupling stability, soliton lifetime, and immunity to environmental noise. The architecture developed in this study lays a strong foundation for future integration into compact modules, paving the way for portable and robust microcomb sources in real-world photonic systems.

# 1. Introduction

Optical frequency combs (OFCs) [1-3] exhibit a highly discrete multispectral structure characterized by a series of evenly spaced optical frequency modes resembling the teeth of a comb. OFCs combine broadband spectral coverage, narrow linewidths, and equidistant frequency spacing, making them an indispensable tool in modern photonics. By leveraging their unique ability to form a frequency-coherent link between optical and electric domains, OFCs have been used as optical frequency rulers calibrated by electric frequency standards or as frequency gears that connect the optical and electric regions. These capabilities have driven advancements in a wide range of fields, including optical frequency metrology [1-3], spectroscopic measurements [4-6], distance ranging [7, 8], imaging [9-11], and sensing [12, 13]. In parallel with these application developments, the advancement of OFC light sources themselves has also progressed. Following Ti:Sapphire-laser-based OFCs [14-16], fiber-laser-based OFCs [17,18], and electro-optic-modulator-based OFCs [19,20], microresonator-based OFCs, or microcombs, have emerged as the new generation of OFCs [21-23]. Microcombs utilize high-Q microresonators composed of highly nonlinear optical waveguides such as silicon nitride ($Si_3N_4$ or SiN), where stable soliton comb generation is achieved by coupling a narrow-linewidth continuous-wave (CW) pump laser light into the microresonator. By harnessing the unique properties of microresonators, microcombs are poised to enable transformative applications such as ultra-stable signal generation of microwave [24], millimeter wave [25], and THz wave [26,27], ranging [28,29], optical communication [30], wireless communication [31,32], and optical computing [33].

Microresonators, which can be fabricated using semiconductor processes, hold great promise as compact and cost-effective OFC sources for future applications, offering versatility that was lacking in earlier OFC platforms. Recent trends in microcomb development include integrated packaging strategies based on self-injection locking and heterodyne coupling with pump lasers [34,35]. However, for typical microresonators with moderate Q-factors (~$10^6$), efficient microcomb generation requires coupling of high-power pump light, amplified by erbium-doped fiber amplifiers (EDFAs), into the microresonator. Currently, coupling pump light into such microresonators often depends on lensed-fiber-based free-space systems, which require bulky microscopes and precision multi-axis stages for alignment [36], posing a major challenge for miniaturization and practical system integration. While such configurations can achieve typical coupling efficiencies on the order of a few tens %, they make the system less practical for compact implementations. Also, free-space coupling using a lensed fiber, which is typically made from a single-mode fiber (SMF), suffers from coupling efficiency degradation due to slow thermal drift of the fiber chip position and/or the pump light polarization. Moreover, the fiber tip is subject to thermal wobbling, which prevents stable high-power excitation required to induce strong nonlinear optical effects and ensure sufficient intracavity optical energy in the microresonator, typically limiting the input pump power through the lensed fiber to around 500 mW. To overcome the limitations of free-space coupling using lensed fibers, several approaches have been reported. For example, a planar lightwave circuit (PLC) mode converter was attached to the input and output ports of the microresonator using optical adhesive after converting the fiber mode profile (~8 µm², typical of conventional SMFs) to match the 4-µm$^2$ mode profile of the inverse taper on the microresonator chip, resulting in a total fiber-chip-fiber coupling loss of less than 6 dB [37]. Another approach involved the use of a spot-size converter for edge coupling a fiber array to the microresonator chip, achieving a total throughput loss of 6.2 dB [38]. Additionally, a high numerical aperture (NA) fiber with a mode field diameter of ~4.1 µm was used to directly mode-match and glue-couple the fiber mode to the inverse taper mode on the chip facet, yielding a fiber-chip-fiber coupling efficiency of 15% [39]. These methods not only enable significant miniaturization of the system but also mitigate coupling efficiency degradation caused by temperature drift. Furthermore, if the optical connection demonstrates sufficient resistance to high optical power, the system could support the use of high-power pump lasers, allowing for stable soliton microcomb generation with higher intracavity optical energy. However, one remaining challenge in these configurations is that they have employed standard SMFs. While SMFs are extremely common, their polarization states are highly sensitive to environmental disturbances such as vibration, strain, and temperature. Since the optical waveguides of microresonators possess polarization dependence, it is necessary to precisely adjust the polarization state using polarization controllers (PCs) to maximize coupling efficiency. These issues could potentially be resolved by directly coupling polarization-maintaining single-mode fibers (PMFs) to the microresonator chip. PMFs offer intrinsic polarization stability, which could eliminate the need for external PCs and improve overall system robustness. To the best of our knowledge, however, no prior studies have demonstrated this direct coupling approach, highlighting an unexplored yet potentially impactful direction. Key technical challenges include aligning the PMF's birefringent axes with the polarization axis of the waveguide mode, and confirming whether comparable coupling efficiency to SMF-based systems can be maintained. This alignment is technically nontrivial, as even slight misalignment of the PMF's birefringent axes can cause polarization rotation, leading to coupling loss and reduced stability.

   Another major challenge is ensuring the reproducibility of stable microcomb operation. For reliable generation and long-term maintenance of soliton microcombs, the relationship between the resonance wavelength of the microresonator and the wavelength of the pump laser is critically important. However, the resonance wavelength of a microresonator is determined by its optical resonator length, which varies due to thermal refractive effects and geometric thermal contraction, making it sensitive to ambient temperature and thus subject to day-to-day fluctuations. As a result, current laboratory-scale systems often require precise daily tuning of the pump laser wavelength to match the thermally shifted resonance wavelength. Looking ahead to the future deployment of microcombs in field applications outside the laboratory, it is crucial to ensure reproducible and stable operation under varying environmental conditions. Prior studies have attempted to address this issue by employing precise temperature control of the microresonator to stabilize its resonance characteristics [40]. Nevertheless, the extent to which temperature control can quantitatively ensure reproducibility of the optical resonance conditions has not yet been fully evaluated. For example, it remains unclear whether temperature stabilization can consistently maintain the resonance frequency within a sub-GHz range, which is typically required for stable soliton microcomb operation.

If these two challenges can be addressed, it will pave the way for the development of compact, robust, and cost-effective microcomb systems, thereby expanding their accessibility for a wide range of practical applications. In this study, we aim to contribute to solving these challenges by utilizing a microcomb system based on the auxiliary laser-assisted thermal balance technique [35,41,42], which utilizes a secondary laser as an auxiliary laser to compensate for the thermo-optic effect and extend the soliton existence range. This method serves as a unified solution to both the coupling and stabilization challenges, supporting robust initiation and long-term maintenance of soliton microcombs. To tackle the first challenge, we directly coupled a high-NA PMF to the microresonator facet using UV-curable optical adhesive. This configuration achieved high coupling efficiency, making it suitable for stable soliton microcomb generation. This method not only retains the advantages of conventional SMF-coupled microresonator systems, such as simplified device configuration and high temporal stability of coupling, but also provides a substantial improvement in robustness against environmental perturbations, thereby eliminating the need for external PCs. To address the second challenge, we mounted the microresonator chip onto a copper block integrated with a thermistor and a Peltier element, enabling precise temperature control. We then evaluated the resonance wavelength reproducibility of the temperature-stabilized microresonator by monitoring the required wavelengths of both the pump laser and the auxiliary laser for initiating soliton microcomb generation. These efforts contribute to the realization of a practical microcomb platform suitable for real-world deployments beyond laboratory environments.

## 2. Experimental setup
### 2.1 Direct connection of high-NA PMF to SiN microresonator using optical adhesive

We used a SiN ring-shaped microresonator (custom, LIGENTEC, S.A., free spectral range = 300 GHz, Q factor ≈ $10^6$). Figure 1 shows a schematic drawing of the direct connection system. A standard PMF (coating diameter = 250 µm, cladding diameter = 125 µm, mode field diameter or MFD = 10.5 µm at 1.55 µm) with an FC/APC connector at one end was connected to a high-NA PMF (coating diameter = 250 µm, cladding diameter = 125 µm, MFD = 4 µm at 1.55 µm) by aligning their MFDs using TEC fusion splicing. The high-NA PMF was directly connected to the input optical waveguide (height = 0.8 µm, width = 1.0 µm) of the SiN microresonator through a single-core fiber array (FA) equipped with a Tempax glass substrate for structural support.

We used a UV-curable optical adhesive to fix the high-NA PMF and the waveguide. Both the input aperture of the optical waveguide and the fiber end face were polished to optical-grade smoothness. To avoid losses caused by refractive index mismatching due to an air gap at the contact interface, a refractive-index-matching liquid was applied between the two end faces to optically connect them without refractive index mismatching. After aligning the high-NA PMF with the input aperture of the waveguide, the optical adhesive was applied. Once the alignment was optimized, UV light was used to cure the adhesive, permanently fixing the fiber to the waveguide interface. This approach ensures reliable coupling with high efficiency and mechanical stability for the direct connection system of the high-NA PMF to SiN microresonator. Using this optical adhesive-based approach, we confirmed that the coupling interface can withstand laser powers of at least 3 W. The output waveguide of the SiN microresonator utilized an optical configuration identical to the input side, comprising a single-core FA, a high-NA PMF, a standard PMF, and an FC/APC connector.

For comparison, we also evaluated two additional coupling configurations: (1) a free-space coupling system using a lensed SMF and (2) a direct-connection system using high-NA SMF. The free-space coupling system was previously detailed elsewhere [36]. The high-NA SMF-based direct-connection system was constructed using the same procedure as the PMF-based setup, but with a standard SMF (coating diameter = 250 μm, cladding diameter = 125 μm, MFD = 10.5 μm at 1.55 μm) and a high-NA SMF (coating diameter = 250 μm, cladding diameter = 125 μm, MFD = 3.2 μm at 1.55 μm) instead of PMFs.

2.2 Temperature control of SiN microresonator

Figure 2 shows a schematic diagram of the temperature control system for the SiN microresonator. The SiN microresonator, connected to a high-NA PMF, is placed on the top surface of a copper block (dimensions: 25 mm height × 25 mm width × 25 mm depth) with thermal grease applied between the two to enhance thermal conductivity. A thermistor (NTC type, resistance = 10 kΩ) is embedded within a hole in the copper block to monitor the temperature near the SiN microresonator. The bottom surface of the copper block is attached to the top surface of a Peltier device (dimensions: 12 mm width × 12 mm depth, $\Delta T_{max}$ = 65.8 °C, $Q_{max}$ = 5.4 W) via thermal grease, which enables the temperature of the SiN microresonator to be precisely adjusted through the thermally conductive copper block. The underside of the Peltier device is attached to an aluminum block, also via thermal grease, which acts as a heat sink. The temperature of the SiN microresonator was maintained at 25.0 ± 0.1 °C using a PID temperature controller (5240 TECSource, 4A/7V, Arroyo Instruments, LLC), with the thermistor serving as the temperature sensor and the Peltier device functioning as the temperature control element.

2.3 Soliton microcomb generation based on auxiliary laser-assisted thermal balance technique

Compared with several methods for the soliton microcomb generation including pump frequency scanning [43], thermal tuning [44], chirped modulation [45], power kicking [46], and self-injected locking [47], the auxiliary laser-assisted thermal balance technique [36,41,42], which involves the combined use of a pump laser and an auxiliary laser, offers advantages in terms of compensation of the thermo-optic effect, tuning-speed independence, and extension of the soliton existence range. Figure 3 shows a schematic diagram of the soliton microcomb generation system. The pump source is a tunable CW laser (TSL-550, Santec, wavelength = 1550–1630 nm, linewidth = 200 kHz, output power = 100 mW), amplified to 480 mW by a high-power EDFA (EDFA1; EFAP1BB12SS07A, GIP Technology Corp., wavelength = 1540–1656 nm, saturated output power = 3 W). The amplified light is routed through an optical circulator (OC1) and coupled into the microresonator via a standard PMF and a high-NA PMF. The auxiliary source is another tunable CW laser (TLX1, Thorlabs, wavelength = 1525–1610 nm, linewidth = 10 kHz, output power = 10 mW), amplified to 500 mW by a second EDFA (EDFA2; laboratory-made, center wavelength = 1550 nm, saturated output power = 0.5 W). After passing through another circulator (OC2), it is injected into the microresonator in the reverse direction via the same PMF pair used for the pump light. The slow and fast axes of the high-NA PMFs on both the pump and auxiliary sides were aligned transversely and longitudinally, respectively, with respect to the waveguide aperture of the microresonator to ensure optimal polarization matching. As a result, the pump and auxiliary lights propagate in counterclockwise and clockwise directions, respectively, within the ring-type microresonator. For the comparative configuration using a high-NA SMF-coupled microresonator, polarization controllers were inserted (not shown) between the optical circulators and the high-NA SMFs to optimize polarization.

To generate a soliton microcomb, the auxiliary light is first set near the shorter-wavelength side of the microresonator resonance. The pump light is then swept from a wavelength shorter than the auxiliary light toward longer wavelengths. As it crosses the auxiliary wavelength, a chaotic comb begins to form. With further red detuning of the pump, the spectrum becomes distorted and exhibits irregular envelope modulation, characteristic of the transition from chaotic to soliton combs. At this point, the auxiliary light is gradually swept toward shorter wavelengths, thereby leading to the formation of a soliton microcomb.

The generated soliton microcomb is monitored using an optical spectrum analyzer and a power meter after passing through OC2. Residual pump components can be removed with an optical bandstop filter (OBSF) if necessary. Alternatively, the soliton microcomb may be optically amplified using an additional EDFA (EDFA3). The auxiliary light power is independently monitored after passing through OC1.

## 3. Results

3.1 Stability evaluation of coupling efficiency in high-NA PMF-based direct connection system

We evaluated the coupling efficiency and its stability in the high-NA PMF-based direct connection system. In this evaluation, only the pump light was used, with its wavelength detuned from the resonant wavelength of the microresonator, while no auxiliary light was injected. At the start of the measurement, the input power of the high-NA PMF on the pump side was 480 mW, while the output power of the high-NA PMF on the opposite side was 277 mW. This resulted in an initial coupling efficiency of 57.7%. To analyze the coupling efficiency stability over time, the output power was normalized to its initial value, and the time-dependent change in the normalized output power was plotted as a percentage relative to the initial value, as shown by the red trace in Fig. 4(a). A magnified view of this fluctuation ratio is presented in Fig. 4(b). The fluctuation ratio of the high-NA PMF-based direct connection system remained nearly constant, with only a minimal decrease of 0.1%. As a result, the coupling efficiency after 10 hours was maintained at 57.7%, consistent with its initial value at the start of the experiment. This high level of stability can be attributed to the robustness of the PMF-based direct connection design.

As a reference, we next evaluated the performance of a conventional lensed-SMF-based free-space coupling system under identical conditions. At the beginning of the measurement, this system exhibited a total loss of 51.7%, with an input power of 480 mW and an output power of 248 mW. The time-dependent normalized output power is shown in the green trace of Figs. 4(a) and 4(b). Unlike the high-NA PMF-based direct connection system, the free-space system showed significant instability, with the fluctuation ratio continuously decreasing over time and eventually reaching a normalized output power decrease of 81%. Consequently, the coupling efficiency at the end of the measurement dropped from the initial value of 51.7% to 9.8%. Such a pronounced reduction in pump power and coupling efficiency makes continuous soliton microcomb operation infeasible. These results clearly highlight the superiority of the high-NA PMF-based direct connection system, demonstrating that it provides an optimal platform for long-term stable operation of soliton microcombs.

As another reference, we also evaluated the performance of the high-NA SMF-based direct connection system under identical conditions. This configuration showed a total loss of 41.5% (input power = 480 mW, output power = 199 mW). Although the high-NA SMF has a smaller MFD and is expected to provide better mode matching with the waveguide aperture of the microresonator, the measured coupling efficiency was lower than that of the PMF-based system. This reduced efficiency may be attributed to the larger beam divergence inherent to high-NA SMFs. The tighter optical confinement leads to a more divergent beam at the output facet, increasing sensitivity to angular and positional misalignments. These factors can result in greater insertion loss and reduced alignment tolerance, ultimately degrading the coupling performance. The blue trace of Figs. 4(a) and 4(b) shows the corresponding time-dependent normalized output power in the high-NA SMF-based direct connection system. A fluctuation of only 1.2% was observed over 10 hours, indicating good stability, and demonstrating the advantages of fiber connection. As a result, the coupling efficiency after 10 hours was maintained at 41.0%, only slightly decreased from the initial value of 41.5%.

From the comparison among the three systems, two key insights emerge. First, in the lensed-SMF-based free-space coupling system, the significant degradation is primarily attributed to slow thermal drift that causes displacement of the fiber-chip position, rather than to thermal drift of the pump light polarization.   Second, in the high-NA SMF-based direct connection system, the minor variation is likely due to temperature-induced drift in the pump light's polarization state. This comparison clearly demonstrates the superiority of the high-NA PMF-based direct connection system in terms of coupling efficiency and long-term stability, making it highly effective for sustaining soliton microcomb operation.

3.2 Repeatability and stability of resonance properties in temperature-controlled microresonator

We first evaluated the resonance repeatability of a temperature-controlled microresonator by generating a soliton microcomb five times and recording the wavelengths of both the pump and auxiliary lasers in each trial. The wavelengths were obtained directly from the set values of the respective laser controllers. The standard deviation of wavelength in the pump and auxiliary lasers was used as a metric for repeatability. Without temperature control, the optical frequency variation of the pump laser was 1.56 GHz, corresponding to a repeatability of $8.07 \times 10^{-6}$. With temperature control, this variation was reduced to 0.92 GHz, improving repeatability to $4.76 \times 10^{-6}$, a 41% reduction in frequency variation. For the auxiliary laser, the optical frequency variation decreased from 10.32 GHz (repeatability = $5.34 \times 10^{-5}$) to 0.86 GHz (repeatability = $4.45 \times 10^{-6}$), corresponding to a 91.7% reduction. Although our laboratory environment is already maintained under air conditioning with a room temperature stability of approximately 19.8 ± 0.3 °C, these results clearly demonstrate the effectiveness of active microresonator temperature control. In environments subject to larger temperature fluctuations, this advantage is expected to be even more significant. Furthermore, temperature control of the microresonator significantly reduces the required tuning range of the optical frequency (or wavelength) for both the pump and auxiliary lasers. As a result, the use of general-purpose laser sources with narrower tunability, such as distributed feedback (DFB) lasers, becomes feasible. Although widely tunable laser sources were used in this study, the enhanced resonance repeatability enabled by temperature stabilization paves the way for the use of more compact and cost-effective laser sources.

In addition, we evaluated the stability of the resonance property in a temperature-controlled microresonator. In this experiment, the robustness of the soliton state against intentional thermal perturbations was assessed by monitoring the power of the generated soliton microcomb. To apply controlled thermal perturbations, a household hair dryer (temperature of warm air = 120 °C, air volume = 0.33 $m^3$/min) was used. A cloth cover was placed over the nozzle of the dryer to prevent direct exposure of the microresonator to heated airflow, thereby minimizing the influence of air turbulence and mechanical vibrations. By varying the distance between the hair dryer and the microresonator while continuously blowing warm air, the degree of heating could be adjusted: 6 cm for mild heating, 4 cm for moderate heating, and 2 cm for strong heating. After confirming soliton step formation by monitoring the microcomb power, the thermal perturbation was initiated. The actual temperature near the microresonator was monitored using a second thermistor, placed just above the microresonator without mechanical contact, and separate from the one used for temperature control. Figure 5(a) shows the time variation of the soliton microcomb power and the corresponding microresonator temperature in the temperature-controlled case over a 200-min measurement period. Before applying the thermal perturbation, the power fluctuation was limited to 82.5 ± 0.4 mW, and the temperature fluctuation remained within 23.3 ± 0.3 °C. After applying the perturbation by stepwise changing the hair dryer distance closer and then farther, the microresonator temperature responded accordingly, showing a stepwise variation within the range of 34 to 69 °C. Despite these large temperature fluctuations, the soliton microcomb power remained nearly constant (81.5 ± 0.8 mW), indicating that the soliton state was maintained. Figure 5(b) presents a magnified view of the power and temperature traces in Fig. 5(a) during the period from 8 to 18 minutes. The fast fluctuation observed in power trace is likely caused by thermo-optic effects induced by the hair dryer heating. For comparison, Fig. 5(c) shows the results for the same perturbation applied to a microresonator without active temperature control over a measurement time of 10 min. When the hair dryer was fixed at a distance of 6 cm, the temperature gradually increased, and the power showed slight variations in response. However, as the temperature approached a rise of nearly 10 °C, the soliton state could no longer be sustained and collapsed.

In the above experiment, the thermal perturbation by the hair dryer was applied after the generation of the soliton microcomb. To further assess the robustness of the temperature-controlled microresonator, we next investigated whether soliton microcomb generation is still achievable under pre-heated conditions. Specifically, the hair dryer was positioned at a distance of 4 cm to apply moderate heating before initiating the soliton generation sequence. Figure 5(d) shows the time variation of the microcomb output power and the corresponding temperature during this 60-min test. When the hair dryer was turned on after 3 minutes, a temperature rise and the following fluctuation similar to that shown in Fig. 5(a) was observed. At approximately 4 minutes, the pump and auxiliary laser wavelengths were tuned according to the same procedure described earlier under this thermal perturbation. The resulting power trace exhibited the characteristic transitions, from chaotic comb to multi-soliton, and finally to a stable soliton step, confirming that single soliton microcomb generation was successfully achieved even under heated conditions. For comparison, we conducted the same experiment using a microresonator without active temperature control. In this case, none of the expected states, chaotic comb or soliton step, were observed (data not shown). This result reinforces the critical role of temperature control in ensuring stable initiation of soliton microcomb operation.

Overall, these findings confirm the critical role of active temperature control in both initiating and maintaining a stable soliton microcomb operation under thermal fluctuations.

3.3 Soliton microcomb generation

We evaluated the consistency and reliability of the spectral characteristics of generated soliton microcombs by comparing three coupling schemes: a high-NA PMF-based direct connection, a lensed-SMF-based free-space coupling, and a high-NA SMF-based direct connection. The optical spectra of the soliton microcombs generated using the high-NA PMF-based direct connection is shown in Fig. 6(a), exhibiting the distinctive spectral envelopes characteristic of soliton microcombs. Specifically, the spectra follow a sech2-shaped distribution, with maximum intensity at the center and a symmetric, gradual decay toward both sides. This profile reflects the temporal shape of soliton pulses, which are mathematically described by the hyperbolic secant (sech) function in the time domain. For comparison, soliton microcombs were also generated using the lensed-SMF-based free-space coupling scheme and the high-NA SMF-based direct connection, as shown in Figs. 6(b) and Fig. 6(c). The resulting spectra similarly exhibited the characteristic sech2-shaped envelope, confirming that both coupling methods can reliably produce equivalent soliton microcombs. These results validate the effectiveness of the high-NA PMF-based direct connection system as a practical and robust alternative to conventional free-space coupling or the high-NA SMF-based direct connection for soliton microcomb generation.

To assess the long-term stability of soliton microcombs, we investigated the duration over which the soliton state could be continuously maintained. Similar to the spectral comparison above, the evaluation was conducted across all three coupling schemes to determine the relative advantages in terms of operational stability. Here, the duration of soliton microcomb operation was investigated using the three coupling schemes. First, a soliton microcomb was generated using the high-NA PMF-based direct connection system. The operation duration exceeded 24 hours, at which point the experiment was terminated due to scheduled system shutdown, suggesting the potential for even longer stable operation. Second, when a lensed SMF-based free-space coupling configuration was used, the operation duration was limited to only 30 minutes. This short duration is likely due to a gradual decrease in coupling efficiency caused by thermal drift and thermal wobbling, which led to insufficient intracavity optical energy to sustain the soliton microcomb. Third, a direct connection using a high-NA SMF was employed. The soliton microcomb was maintained for a comparably long duration (> 24 hour), similar to that achieved with the high-NA PMF-based system. This result is consistent with the long-term stability of the coupling efficiency confirmed in Fig. 4 for the high-NA SMF-based direct connection. Although a slight degradation in coupling efficiency was observed due to thermally induced polarization changes, it had minimal impact on the stability of soliton microcomb operation, indicating the robustness of this coupling scheme.

To further highlight the advantage of the high-NA PMF-based direct connection over the high-NA SMF-based counterpart, we evaluated the operational robustness of the soliton microcomb under intentional environmental disturbances applied to the fiber sections of each coupling system. While the experiments shown in Fig. 5 focused on external thermal perturbations applied directly to the microresonator, the present test targeted the fiber sections, which are more susceptible to external influences due to their physical length and exposure. Three types of disturbances were introduced: (i) mechanical disturbance by slight fiber bending, (ii) thermal disturbance by applying hot air and then removing it, and (iii) sequential thermal disturbance by applying hot air followed by cold air. The comparative results are summarized in Table 1. In the case of the high-NA PMF-based direct connection, the soliton microcomb remained stable under all disturbance conditions. In contrast, the high-NA SMF-based direct connection failed to sustain the soliton state in every case. This difference is likely due to the contrasting environmental robustness of PMF and SMF. PMFs maintain the excitation conditions of the microresonator more stably under perturbation by suppressing fluctuations in the polarization of the pump light, whereas SMFs are more prone to polarization changes that lead to reduced coupling efficiency and ultimately prevent the soliton state from being sustained. These results clearly underscore the superior stability of the high-NA PMF-based direct connection and suggest that it offers significant advantages for future applications of microcombs in environments subject to various external disturbances outside the laboratory.

## 4. Discussion

One of the key contributions of this study is the realization of a compact and robust architecture for soliton microcomb systems. By combining high-NA PMFs for coupling with standard PMFs for delivery, the system achieves excellent environmental robustness and enables long-duration soliton operation without alignment drift. Direct fiber coupling also simplifies the optical setup, eliminating the need for microscopes or multi-axis positioning stages. Additionally, active temperature control significantly improves reproducibility, reducing the optical frequency variation by approximately 41% for the pump laser and 91.7% for the auxiliary laser. These improvements provide a solid foundation for stable and repeatable soliton microcomb generation in real-world, non-laboratory environments. The comparative evaluation of three coupling schemes, PMF-based direct connection, SMF-based direct connection, and lensed-SMF-based free-space coupling, offers clear design guidelines for future soliton microcomb platforms. The PMF-based configuration demonstrated superior coupling efficiency, stability, and soliton lifetime, particularly under thermal, airflow, and mechanical disturbances. These characteristics, when combined with existing compact photonic packaging technologies [48], are expected to further enhance the practical usability of soliton microcombs. The approach presented here provides a compelling pathway toward highly deployable and versatile microcomb systems, particularly in the form of integrated microresonator modules that can serve as plug-and-play components in broader photonic platforms.

Another promising direction suggested by this work is the realization of portable microcomb light sources based on integrated microresonator modules. To achieve this goal, however, several practical challenges remain to be addressed. The first challenge lies in the choice of laser sources. While the temperature control introduced in this study effectively relaxes the requirements on laser tunability, enabling the use of general-purpose DFB lasers instead of expensive and bulky external cavity laser diodes (ECLDs), the current architecture still relies on a two-laser configuration utilizing an auxiliary laser-assisted thermal balance technique. Although this method offers advantages such as enhanced thermal stability and expanded soliton existence range, the requirement for two separate laser sources poses a barrier to system miniaturization and integration. Fortunately, various methods for generating soliton microcombs with a single laser have been proposed, including power kicking [46] and laser self-injection locking [47]. The architecture presented in this work is inherently compatible with such approaches, suggesting a feasible pathway toward simplifying the laser source subsystem.

In addition to laser source simplification, another important consideration is the reliance on EDFAs. While EDFAs are essential in the present setup to compensate for the inherently low efficiency of converting pump light into a microcomb, typically less than 1% in the case of a microresonator with $Q \approx 10^6$, they hinder miniaturization, increase cost, and introduce additional power consumption factors that are incompatible with portable or battery-operated devices. Moreover, EDFAs act as significant heat sources within a compact enclosure, potentially giving rise to thermal management issues that further complicate packaging and long-term operational stability. Alternatively, under the assumption that current Q-factor microresonators remain in use, integrated waveguide-type EDFAs [49] may serve as a more compact and power-efficient amplification solution.

In parallel, further advances in high-Q resonator fabrication, especially targeting Q-factors exceeding $10^7$, would help eliminate the need for EDFAs altogether by enabling sufficient microcomb generation efficiency without external amplification.

## 5. Conclusion

In this work, we presented a compact and robust architecture for soliton microcomb generation based on a direct fiber connection between high-NA PMFs and SiN microresonators, complemented by active temperature control. This configuration eliminates the need for free-space optics or complex multi-axis alignment stages, significantly simplifying the system setup. The high-NA PMF-based connection demonstrated excellent coupling efficiency and long-term stability, enabling soliton microcomb operation for over 24 hours without interruption. Additionally, active temperature control significantly improves reproducibility, reducing the optical frequency variation by approximately 41% for the pump laser and 91.7% for the auxiliary laser. A quantitative comparison among three coupling schemes, PMF-based direct connection, SMF-based direct connection, and lensed-SMF-based free-space coupling, confirmed the superior performance of the proposed system across all relevant metrics, including coupling stability, soliton duration, and resilience to thermal, mechanical, and airflow perturbations.

These results suggest a promising pathway toward deployable and field-ready microcomb systems. The demonstrated robustness against environmental disturbances and the compatibility with compact packaging technologies make the proposed architecture particularly suitable for integration into photonic modules. Such integrated microresonator modules could serve as plug-and-play components in a variety of photonic platforms, supporting applications in portable frequency metrology, broadband spectroscopy, optical sensing, and high-capacity communication systems. The simplified alignment-free architecture not only enhances reliability but also paves the way for the practical implementation of soliton microcombs outside controlled laboratory environments.

To further advance toward fully integrated microcomb light sources, several practical challenges remain. The current system architecture relies on a two-laser configuration with an auxiliary laser-assisted thermal balance technique, which, while effective, adds complexity and limits integration. The use of temperature control alleviates the requirement for widely tunable lasers and allows the potential adoption of general-purpose DFB lasers. Furthermore, replacing the two-laser scheme with single-laser excitation techniques, such as power kicking [46] and laser self-injection locking [47], could further simplify the system. Another important issue is the reliance on EDFAs, which are bulky, power-hungry, and generate significant heat that complicates thermal management in enclosed systems. While integrated waveguide-type EDFAs offer a potential interim solution for low-Q microresonators, a more sustainable long-term strategy involves developing high-Q resonators ($Q > 10^7$) that support efficient microcomb generation without external amplification. These combined efforts will be essential to realize fully integrated, power-efficient, and portable soliton microcomb sources for real-world deployment.

**Funding**

Ministry of Internal Affairs and Communications of Japan (JPMI240910001, JPJ000254); Cabinet Office, Government of Japan (Promotion of Regional Industries and Universities); Japan Society for the Promotion of Science (Regional Innovation and Excellence, J-PEAKS); Tokushima Prefecture, Japan (Creation and Application of Next-Generation Photonics).

**Acknowledgment**

We would like to express our sincere gratitude to Drs. Shuichiro Asakawa and Naomi Kawakami of NTT Advanced Technology Corporation for their support regarding the PMF-based direct connection and SMF-based direct connection systems of microcavity.

**Data availability**

Data underlying the results presented in this paper are not publicly available at this time but may be obtained from the authors upon reasonable request.

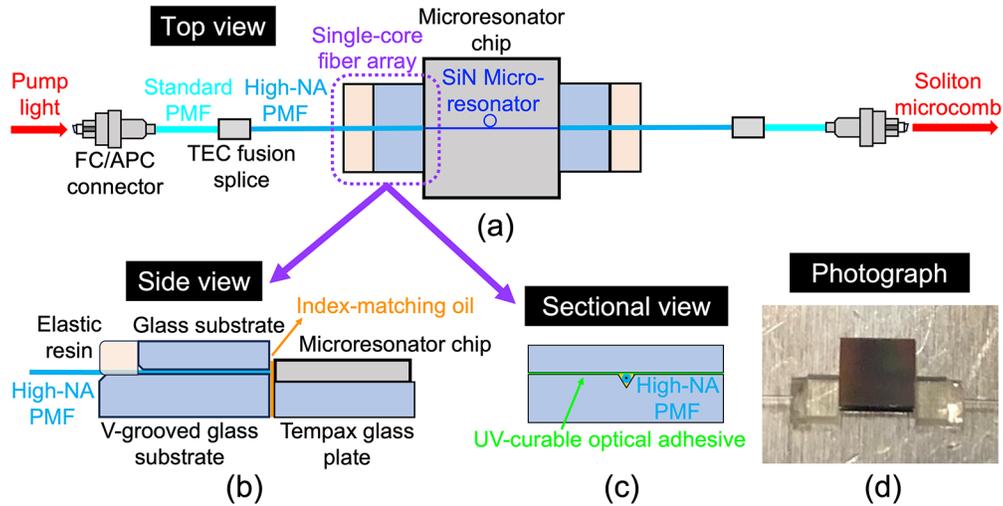

Fig. 1. Schematic illustration and photographs of the fiber-coupled microresonator system using a high-numerical-aperture polarization-maintaining fiber (high-NA PMF). (a) Top view of the direct connection between the high-NA PMF and the SiN microresonator chip via a single-core fiber array (FA). (b) Side view of the FA-based direct fiber connection, showing the FA structure composed of a glass substrate, V-grooved glass substrate, and Tempax glass substrate, with index-matching oil applied between the fiber endface and waveguide input. (c) Sectional view of the FA–microresonator interface, illustrating the UV-curable optical adhesive layer between the glass substrate and V-grooved substrate. (d) Photograph of the fabricated fiber-coupled SiN microresonator system.

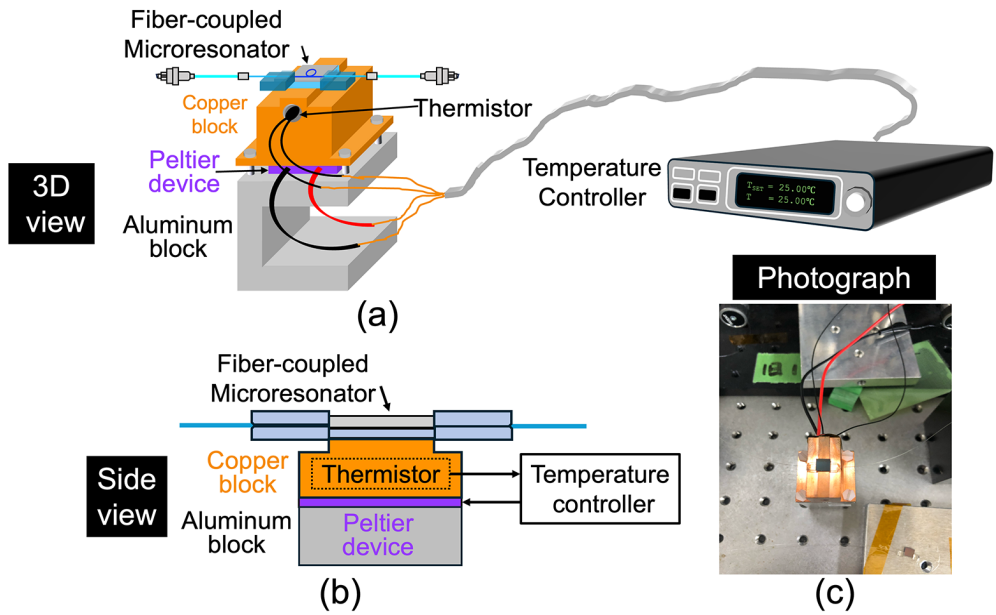

Fig. 2. Configuration of the temperature control system for the SiN microresonator. (a) 3D view and (b) side view of the temperature-controlled copper block mounting the fiber-coupled microresonator. The microresonator chip is thermally stabilized using a thermistor, Peltier device, and aluminum heat sink. (c) Photograph of the assembled temperature-controlled fiber-coupled microresonator module.

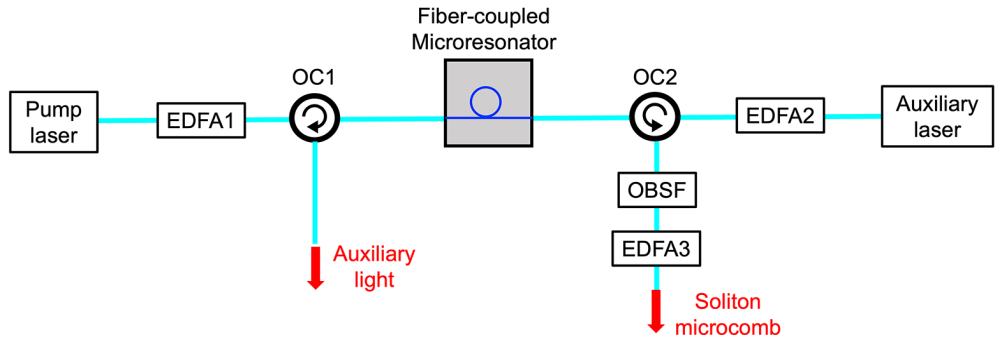

Fig. 3. Experimental setup for soliton microcomb generation using the auxiliary laser-assisted thermal balance technique. Continuous-wave (CW) pump and auxiliary lasers are amplified by erbium-doped fiber amplifiers (EDFA1 and EDFA2) and injected into the fiber-coupled SiN microresonator via optical circulators (OC1 and OC2). An optical bandstop filter (OBSF) is optionally used to suppress residual pump components in the soliton microcomb output. Additionally, a supplementary erbium-doped fiber amplifier (EDFA3) is used to provide optional optical amplification after pump suppression, if necessary.

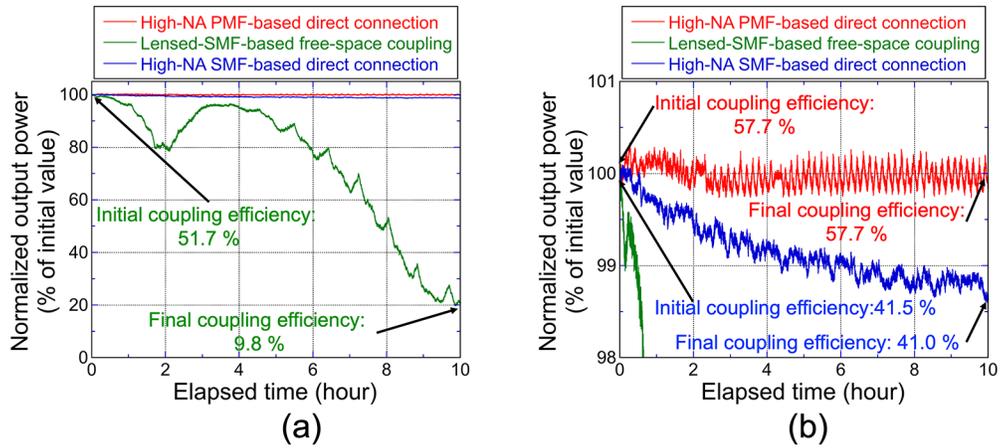

Fig. 4. Comparison of long-term coupling stability among three coupling schemes. (a) Time variation of normalized output power (expressed as a percentage of the initial value) over 10 hours. (b) Same data plotted on a logarithmic scale for better visualization of small fluctuations. Red trace: high-NA PMF-based direct connection system. Green trace: high-NA SMF-based direct connection system. Blue trace: lensed-SMF-based free-space coupling system.

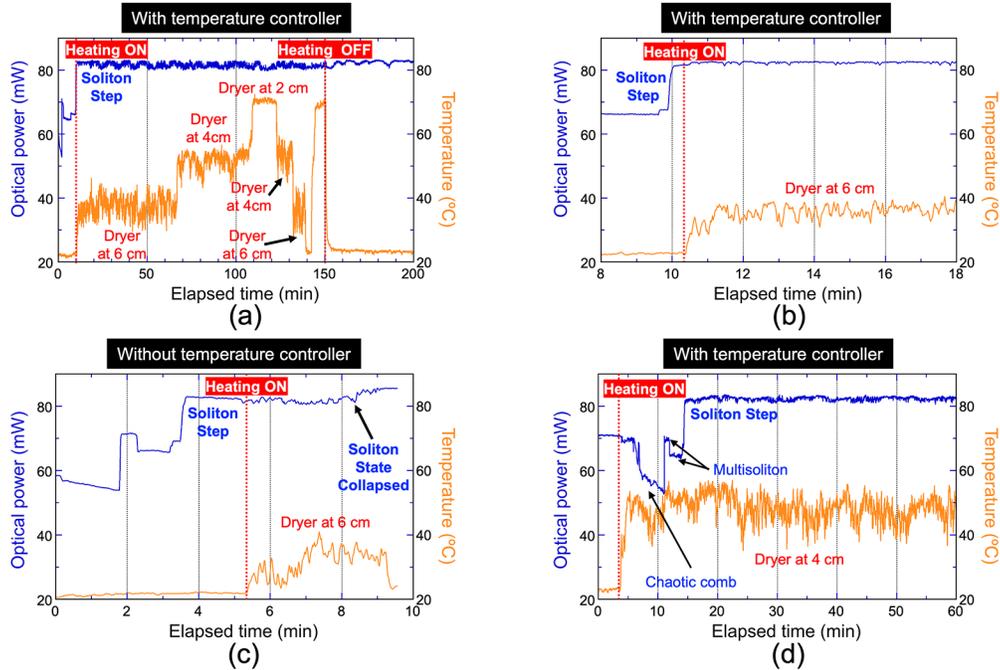

Figure 5. Temperature response and soliton microcomb output power stability under intentional thermal perturbations applied using a hair dryer operated in warm-air mode. (a) Time variation of soliton microcomb output power (blue) and microresonator temperature (orange) under active temperature control. The distance between the hair dryer and the microresonator was varied stepwise to modulate heating intensity. (b) Enlarged view of the stable soliton state in panel (a), highlighting short-term fluctuations caused by dryer-induced perturbations. (c) Soliton microcomb power and temperature traces without temperature control. Collapse of the soliton state was observed following temperature rise due to continuous heating. (d) Soliton initiation and stable operation under pre-heated conditions with active temperature control. Temporal evolution shows successful transition from chaotic comb to multisoliton state and final soliton step. In all cases, a cloth cover was placed over the nozzle of the hair dryer to block airflow, allowing thermal perturbation without introducing mechanical disturbances from air currents.

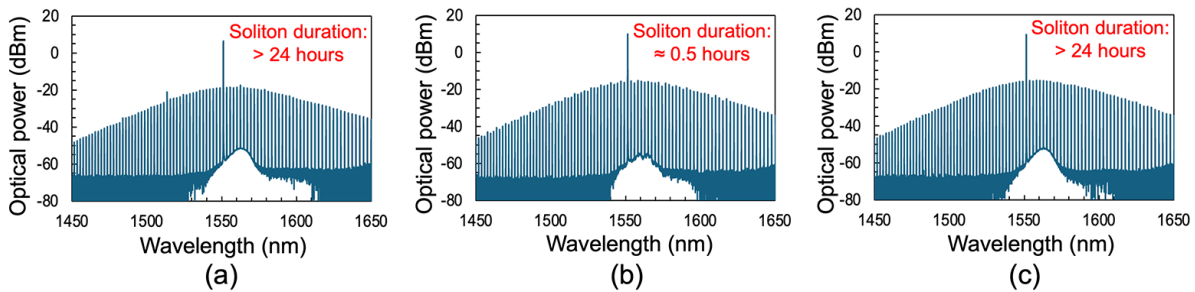

Figure 6. Optical spectra of soliton microcombs generated using three different coupling configurations, along with corresponding soliton operation durations. (a) High-NA PMF-based direct connection: Soliton duration > 24 hours (highlighted in red). (b) Lensed-SMF-based free-space coupling: Soliton duration ≈ 0.5 hours (highlighted in red). (c) High-NA SMF-based direct connection: Soliton duration > 24 hours (highlighted in red). In all cases,

the sech²-shaped spectral envelope characteristic of soliton microcombs is clearly observed; however, significant differences in soliton operation duration were confirmed among the three configurations.

**Table 1. Robustness comparison under external disturbances applied to the fiber sections of the coupling systems.**

|  | Slight Bending | Hot air then removed | Hot air then cold air |
|---|---|---|---|
| PMF-based direct connection | Maintained | Maintained | Maintained |
| SMF-based direct connection | Collapsed | Collapsed | Collapsed |